\documentclass[aps,pra,superscriptaddress,longbibliography,amsmath,amssymb,reprint]{revtex4-2}
\usepackage[T1]{fontenc}
\usepackage{graphicx}
\usepackage{bm}
\usepackage{hyperref}
\hypersetup{colorlinks=true, linkcolor=blue, citecolor=blue, filecolor=cyan, urlcolor=cyan}
\begin{document}

\title{Optical trapping of neutral atoms via magnetic-dipole transitions}

\author{E. Orellana-Paillaleo}
\email{eorellana2020@udec.cl}
\affiliation{Departamento de Física, Facultad de Ciencias Físicas y Matemáticas, Universidad de Concepción, Concepción, Chile}

\author{N. Vera}
\affiliation{Departamento de Física, Facultad de Ciencias Físicas y Matemáticas, Universidad de Concepción, Concepción, Chile}

\author{P. Solano}
\email{psolano@udec.cl}
\affiliation{Departamento de Física, Facultad de Ciencias Físicas y Matemáticas, Universidad de Concepción, Concepción, Chile}

\begin{abstract}
Optical trapping of neutral atoms is almost universally described as an electric-dipole (E1) effect. We show that the magnetic field of light, acting on a magnetic-dipole (M1) transition, can induce an ac Zeeman shift significantly larger than the ac Stark shift, thereby enabling magnetic trapping. We construct the M1 polarizability tensor and identify a critical detuning $\delta_c$, independent of optical power, below which the magnetic interaction dominates. We analyze suitable trapping parameters and possible loss mechanisms, particularly for the intraconfigurational transitions $^3H_6\rightarrow{}^3H_5$ of erbium at $1437.13$~nm and $^5I_8\rightarrow{}^5I_7$ of dysprosium at $2418.84$~nm. We propose two trapping configurations: a hybrid magnetic/electric optical tweezer and an optical lattice with separated magnetic and electric antinodes, producing state-dependent purely magnetic and purely electric trapping wells. Our results present optical magnetic fields as a suitable resource to expand the toolbox for atomic manipulation.
\end{abstract}

\maketitle

\section{Introduction}
\label{sec:introduction}

Optical trapping of neutral atoms, along with micro- and nanoparticles, has traditionally been studied through the interaction between the electric field of light and the induced electric dipole moment \cite{Ashkin1986,Chu1986,Grimm2000}. The magnetic field of light has only recently been used to implement optical forces on high-index silicon nanoparticles \cite{Li2025OptMagTrap}, opening avenues for further extension to atomic manipulation. Although magnetic dipole interactions at optical frequencies are relevant for studying highly forbidden spectroscopic transitions \cite{Shortley1940,Sobelman1979,Sukachev2016Tm,King2022}, the magnetic component of light is often neglected in the context of optical forces for atom trapping. Addressing this oversight could enhance the methodologies available for atomic manipulation, with potential applications in quantum simulation.

The magnetic component of radiation has been utilized to trap \cite{Spreeuw1994} and push \cite{Fancher2018ACZ} neutral atoms exclusively at microwave frequencies via hyperfine magnetic-dipole (M1) transitions, thereby limiting the spatial resolution of atomic control. In contrast, at optical frequencies, the M1 coupling of neutral atoms to the trapping field has been regarded merely as a correction to the frequency of lattice clocks, pertinent to their level of precision \cite{Taichenachev2008}. To the best of our knowledge, no optical trap for neutral atoms has been proposed in which the primary forces originate from the magnetic field of the laser. To address this gap, it is crucial to determine whether a genuine optical M1 transition in a neutral atom can generate a trapping potential that effectively competes with the conventional off-resonant electric-dipole (E1) transitions. The primary challenge is to identify atomic species that exhibit a significant M1 transition at optical wavelengths while maintaining a sufficiently weak response to the E1 transitions.

The electronic structure of lanthanide atoms is defined by localized $4f$ electrons, which allow M1 transitions within the same partially filled orbital. These transitions are distinctly separated from the E1 transitions, which are mainly associated with the closed $6s^2$ shell. The M1 transition occurs in the near-infrared region, significantly distant from the dominant E1 $6s\rightarrow6p$ resonances near $400$~nm. Although the matrix elements and radiative lifetime of the M1 transition have not been measured for atoms in free space, the levels involved are nearly pure Russell-Saunders (LS) states \cite{Kramida2024NIST}, which allows their analytic calculation. Regarding the state-of-the-art manipulation of lanthanides, laser cooling and Bose-Einstein condensation of erbium and dysprosium have established these species as prominent platforms for optical trapping and studying dipolar quantum gases \cite{Lu2010DyMOT,Frisch2012ErMOT,Aikawa2012ErBEC,Lu2011DyBEC,Chomaz2023Review}.

In this work, we demonstrate that the magnetic field of light can successfully dominate optical trapping in neutral erbium and dysprosium atoms. We study the ${}^3H_6\rightarrow{}^3H_5$ transition of Er at $1437.13$~nm and the ${}^5I_8\rightarrow{}^5I_7$ transition of Dy at $2418.84$~nm. Both transitions connect fine-structure levels within the electronic ground configuration \cite{Martin1978,Kramida2024NIST} and combine appreciable M1 coupling with natural linewidths at the hertz level. We identify a critical detuning $\delta_c$, independent of optical power, below which the optical forces on the M1 transition dominate over E1. We demonstrate that operating at an appropriate detuning results in useful trap depths determined by magnetic interactions. Additionally, we propose two experimentally viable configurations: a magnetic optical tweezer and a standing-wave lattice. The latter configuration spatially displaces the electric and magnetic antinodes to create purely magnetic wells at the nodes of the electric field. These isolated magnetic potentials can stably hold spin-polarized atoms for extended periods, providing a robust mechanism for state-dependent quantum control.

\begin{figure*}[t]
\centering
\includegraphics[width=\textwidth]{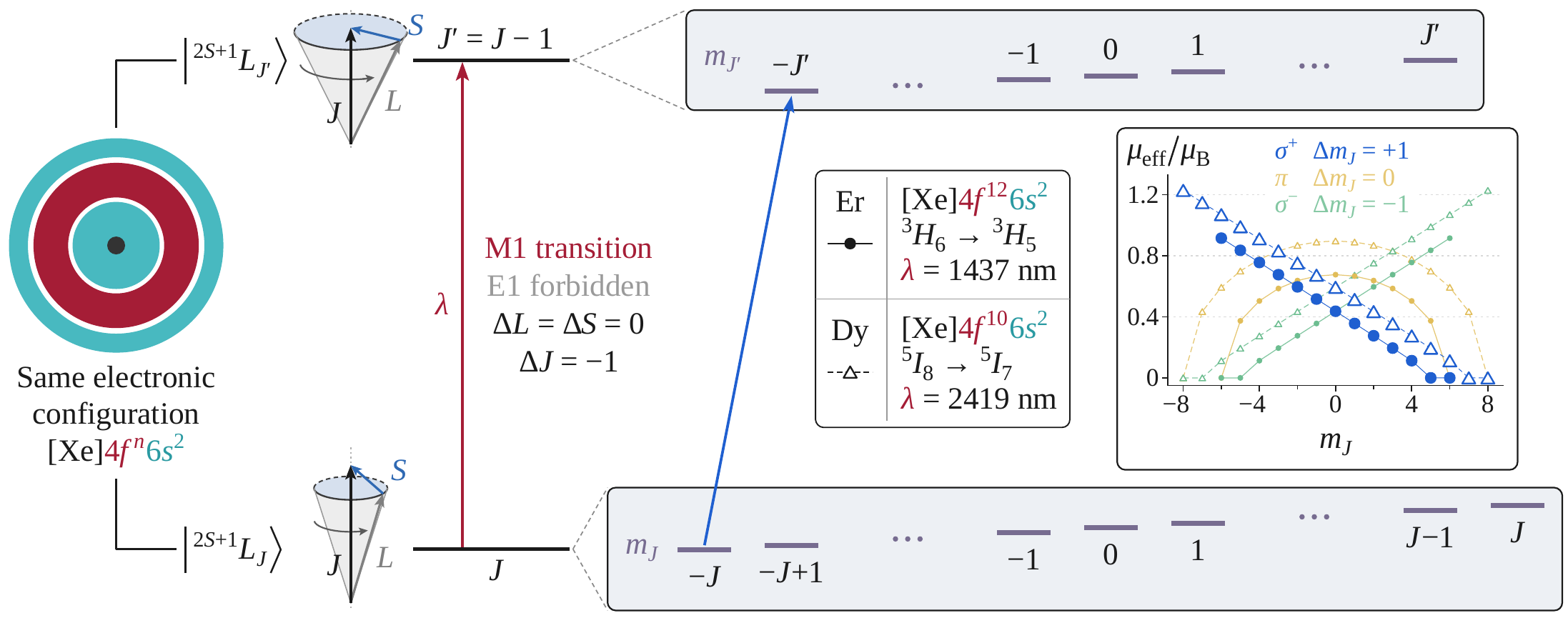}
\caption{Magnetic-dipole transition between the two lowest levels of the ground fine-structure multiplet of erbium and dysprosium. The cones show the vector model of each level: the two levels differ only in the relative orientation of $\bm{L}$ and $\bm{S}$, and the component of $\bm{\mu}\propto\bm{L}+g_s\bm{S}$ perpendicular to $\bm{J}$ couples them. Since both levels belong to the same electronic configuration, and therefore have the same parity, the transition is electric-dipole forbidden. The boxes show the Zeeman sublevels of the bosonic isotopes ($I=0$), and the arrow marks the $\sigma^+$ transition out of the stretched state. The inset shows $\mu_{\rm eff}(m_J)$ for $\sigma^+$ light, Eq.~\eqref{eq:mueff_def}, and, for completeness, for $\pi$ and $\sigma^-$ polarizations.}
\label{fig:transition}
\end{figure*}

\section{Polarizabilities, light shifts, and atomic structure}
\label{sec:atomic}

The interaction between an atom and a classical electromagnetic field can be expanded in multipoles of the atomic charge and current distributions,
\begin{equation}
\hat{H}_{\rm int} = -\hat{\bm d}\cdot\bm E(\bm r,t) - \hat{\bm\mu}\cdot\bm B(\bm r,t) + \mathcal{O}(E2,M2,\ldots),
\label{eq:Hint}
\end{equation}
where $\hat{\bm d}$ and $\hat{\bm\mu}$ are the electric- and magnetic-dipole operators \cite{Sobelman1979,Cowan1981}. Higher multipoles are suppressed by powers of $ka_0=2\pi a_0/\lambda\lesssim2.3\times10^{-4}$ \cite{Sobelman1979}. Relative to the M1 amplitude, the electric-quadrupole (E2) amplitude of the same transition is at most of order $ka_0\langle r^2\rangle_{4f}/(\alpha_{\rm fs}a_0^2)\lesssim3\%$. Hence, we consider only the E1 and M1 interactions, which at second order produce the ac Stark and ac Zeeman shifts, whose spatial variation acts as the trapping potential.

The light shift of a sublevel $|J,m_J\rangle$ is conventionally decomposed into scalar, vector, and tensor contributions, whose relative weights depend on $m_J$, on the light polarization, and on its orientation with respect to the quantization axis \cite{LeKien2013,Lepers2014Er,Becher2018Er,Embacher2024}. The shift produced by a field of amplitude $\bm F$ and polarization $\bm u$ reads
\begin{multline}
U=-\frac{|\bm F|^2}{4}\Biggl[\tilde\alpha_s+\mathcal{A}\cos\theta_k\,\frac{m_J}{2J}\,\tilde\alpha_v \\
+\frac{3\cos^2\theta_p-1}{2}\,\frac{3m_J^2-J(J+1)}{J(2J-1)}\,\tilde\alpha_t\Biggr],
\label{eq:U_general}
\end{multline}
where $\mathcal{A}=i(\bm u^*\times\bm u)\cdot\hat{\bm k}$ is the ellipticity of the light, $\theta_k$ and $\theta_p$ are the angles that the wave vector and the polarization make with the quantization axis, and the three coefficients follow from the coupled polarizabilities $\tilde\alpha^{(K)}$, $K=0,1,2$,
\begin{equation}
\begin{aligned}
\tilde\alpha_s&=-\frac{\tilde\alpha^{(0)}}{\sqrt{3(2J+1)}},\\
\tilde\alpha_v&=\sqrt{\frac{2J}{(J+1)(2J+1)}}\,\tilde\alpha^{(1)},\\
\tilde\alpha_t&=\sqrt{\frac{2J(2J-1)}{3(J+1)(2J+1)(2J+3)}}\,\tilde\alpha^{(2)}.
\end{aligned}
\label{eq:alpha_coeffs}
\end{equation}
In second-order perturbation theory, each $\tilde\alpha^{(K)}$ is a sum over the levels $J'$ that a dipole operator $\hat{\bm V}$ connects to the ground level. Each level contributes its line strength, an angular factor given by a $6j$ symbol, and a resonant denominator,
\begin{multline}
\tilde\alpha^{(K)}=\sqrt{2K+1}\sum_{J'}(-1)^{J+J'}\begin{Bmatrix}1&1&K\\J&J&J'\end{Bmatrix}\bigl|\langle J'\Vert\hat{\bm V}\Vert J\rangle\bigr|^2 \\
\times\frac{1}{\hbar}\,\mathrm{Re}\left[\frac{(-1)^K}{\omega_{J'J}-\omega-i\Gamma_{J'}/2}+\frac{1}{\omega_{J'J}+\omega-i\Gamma_{J'}/2}\right],
\label{eq:alpha_coupled}
\end{multline}
with $\omega_{J'J}$ the transition frequency and $\Gamma_{J'}$ the width of the upper level. Equations~\eqref{eq:U_general}-\eqref{eq:alpha_coupled} require only that $\hat{\bm V}$ be a rank-1 operator contracted with the field. They are usually applied to the E1 interaction, $\hat{\bm V}=\hat{\bm d}$ and $\bm F=\bm E$, and hold without modification for the M1 interaction with
\begin{equation}
\hat{\bm V}=\hat{\bm\mu},\qquad \bm F=\bm B,\qquad \bigl|\langle J'\Vert\hat{\bm\mu}\Vert J\rangle\bigr|^2=S(M1).
\label{eq:dual_substitution}
\end{equation}

\subsection{Magnetic polarizability and ac Zeeman shift}
\label{sec:atomic_polM}
\label{sec:atomic_UM}

The magnetic-dipole operator is $\hat{\bm\mu}=-\mu_B(\hat{\bm L}+g_s\hat{\bm S})/\hbar=-\mu_B[\hat{\bm J}+(g_s-1)\hat{\bm S}]/\hbar$, with $g_s=2.00232$. In LS coupling, because $\hat{\bm J}$ commutes with $\hat{\bm J}^2$, it cannot change $J$, and any M1 transition between levels of different $J$ arises from the spin alone (Fig.~\ref{fig:transition}). Because $\hat{\bm\mu}$ does not act on the radial coordinates, this matrix element involves no radial integral and depends only on $L$, $S$, and $J$. Moreover, $\hat{\bm\mu}$ changes $J$ by at most one unit. In Er and Dy, whose $4f$ shell is more than half filled, the spin-orbit interaction orders the levels of the ground multiplet such that the lowest one has the largest $J$, $J=L+S$, and its only M1 partner is therefore the level $J'=J-1$. The entire magnetic response is thus concentrated in a single transition, whose line strength $S(M1)=|\langle J'\Vert\hat{\bm\mu}\Vert J\rangle|^2$ takes the closed form \cite{Shortley1940,Cowan1981}
\begin{equation}
S(M1)=\frac{(g_s-1)^2\mu_B^2}{4J}\,[(L+S+1)^2-J^2]\,[J^2-(L-S)^2].
\label{eq:S_closed}
\end{equation}

For the transitions considered here, Eq.~\eqref{eq:alpha_coupled} simplifies under four approximations. The sum over $J'$ reduces to the single level $J'=J-1$. The counter-rotating term is negligible, since $|\delta|/\omega_0\sim10^{-6}$ with $\delta=\omega-\omega_0$. Since we work in the regime $|\delta|\gg\Gamma$, only the dispersive part contributes. Finally, at low intensities, where the excited level remains nearly empty, the second-order description is accurate (Sec.~\ref{sec:theory_saturation}). With these approximations, Eq.~\eqref{eq:alpha_coupled} reduces to a single resonant term,
\begin{equation}
\tilde\alpha^{(K)}=\sqrt{2K+1}\,(-1)^{J+J'}(-1)^K\begin{Bmatrix}1&1&K\\J&J&J'\end{Bmatrix}\frac{S(M1)}{\hbar(\omega_0-\omega)}.
\label{eq:alpha_coupled_M1}
\end{equation}
Inserting it into Eq.~\eqref{eq:alpha_coeffs} and evaluating the $6j$ symbols yields
\begin{equation}
\alpha_v(\delta) = \frac{S(M1)}{(2J+1)\hbar\delta}, \qquad \alpha_s=\frac{\alpha_v}{3}, \qquad \alpha_t=-\frac{\alpha_v}{3},
\label{eq:alphaMK_closed}
\end{equation}
where, for consistency with the detuning $\delta$ used throughout, we define $\alpha_s=-\tilde\alpha_s$, $\alpha_v=\tilde\alpha_v$, and $\alpha_t=-\tilde\alpha_t$. For $\sigma^+$ light (Fig.~\ref{fig:transition}), $\mathcal A\cos\theta_k=1$ and $\cos\theta_p=0$. Inserting Eq.~\eqref{eq:alphaMK_closed} into Eq.~\eqref{eq:U_general} shows that the scalar, vector, and tensor parts add up to a single magnetic polarizability of the sublevel, $\alpha_v(\delta)\,r(m_J)$, and that the optical ac Zeeman shift takes the form of a two-level system,
\begin{equation}
\begin{gathered}
U_M(m_J,\bm r)=\frac{|\bm B(\bm r)|^2}{4}\,\alpha_v(\delta)\,r(m_J)=\frac{\hbar\Omega_{m_J}^2(\bm r)}{4\delta},
\end{gathered}
\label{eq:UM_master}
\end{equation}
where $\Omega_{m_J}=\mu_{\rm eff}(m_J)|\bm B|/\hbar$ is the Rabi frequency for the M1 transition. Here, $\mu_{\rm eff}(m_J)$ is the matrix element of the $\sigma^+$ transition from $m_J$, the magnetic analog of the transition dipole moment. The transition is the same for all sublevels, but each couples with a different angular factor, which depends only on its projection $m_J$ onto the quantization axis. Then, the effective magnetic dipole moments read
\begin{equation}
\begin{gathered}
\mu_{\rm eff}(m_J)^2=\frac{S(M1)}{2J+1}\,r(m_J),\\
r(m_J)=\frac{(J-m_J)(J-m_J-1)}{2J(2J-1)}.
\end{gathered}
\label{eq:mueff_def}
\end{equation}

The potential in Eq.~\eqref{eq:UM_master} scales as $1/\delta$ and is attractive for red detuning. The factor $r(m_J)$ equals unity at $m_J=-J$ and decreases to zero at $m_J=J-1$ and $J$, which have no $\sigma^+$ partner in the upper level. We consider the stretched state $m_J=-J$, which has the largest coupling (Table~\ref{tab:atomic}), and is also the lowest Zeeman sublevel, so that fully polarized atoms cannot undergo exothermic dipolar relaxation (Sec.~\ref{sec:loss_collisions}).

\subsection{Electronic configuration and LS purity}
\label{sec:atomic_LS}

Erbium and dysprosium have ground-state configurations $[\mathrm{Xe}]\,4f^{12}6s^2$ and $[\mathrm{Xe}]\,4f^{10}6s^2$, with ground states $^3H_6$ and $^5I_8$. We consider bosonic isotopes, such as ${}^{166}$Er and ${}^{164}$Dy, which have nuclear spin $I=0$ and no hyperfine structure. In both atoms, the electrostatic repulsion among the localized $4f$ electrons fixes the total orbital and spin angular momenta, $L$ and $S$. The spin-orbit interaction then splits the levels with the same $L$ and $S$, which form a fine-structure multiplet, into levels of different total angular momentum $J$, such as ${}^3H_6$, ${}^3H_5$, and ${}^3H_4$ in Er. This hierarchy makes the LS description well suited to the low-lying levels, although the lanthanides do not reach the strict LS limit \cite{Cowan1981,Lepers2014Er,Chomaz2023Review}. The transitions considered here connect two levels of the same $L$ and $S$, with $J\rightarrow J-1$ (Fig.~\ref{fig:transition}), and both levels are nearly pure LS states \cite{Kramida2024NIST,Li2017Dy} whose measured Land\'e factors agree with the LS values (Table~\ref{tab:atomic}). We therefore evaluate the M1 matrix elements in LS coupling.

The same description fixes the natural linewidth of the upper level. Because $\hat{\bm\mu}$ cannot change $L$ or $S$ either, the upper level can only decay by M1 to levels of its own multiplet, and the only one below it is the ground level. In Er, the ${}^3F_4$ level, $1923$~cm$^{-1}$ lower, has a different $L$, so the decay to it is allowed only by the small deviations from LS coupling, and it is further suppressed by the cube of the ratio of the transition energies, $(1923/6958)^3\simeq0.02$; in Dy, there is no other level below. The natural linewidth therefore equals the Einstein coefficient of the M1 transition (Table~\ref{tab:atomic}) \cite{Shortley1940,NISTCompendium},
\begin{equation}
\Gamma = A(M1) = \frac{\mu_0\omega^3}{3\pi\hbar c^3} \frac{S(M1)}{2J'+1}, \qquad \omega=\frac{2\pi c}{\lambda_0}.
\label{eq:AM1}
\end{equation}

\begin{table}[!htb]
\caption{Parameters of the M1 transitions. The Land\'e factors are measured values \cite{Martin1978,Kramida2024NIST}, with the LS values in parentheses, and the LS purities of the lower and upper levels are those reported in Refs.~\cite{Kramida2024NIST,Li2017Dy}. $S(M1)$, $\Gamma$, and $\mu_{\rm eff}$ (for $m_J=-J$ and $\sigma^+$ light) are calculated in LS coupling with the level energies of Ref.~\cite{Kramida2024NIST}.}
\label{tab:atomic}
\begin{ruledtabular}
\begin{tabular}{lcc}
 & Er ($^3H_6 \rightarrow {}^3H_5$) & Dy ($^5I_8 \rightarrow {}^5I_7$) \\
\colrule
$\lambda_0$ (nm) & 1437.13 & 2418.84 \\
$g_J$ & 1.16381 (1.16705) & 1.24159 (1.25058) \\
$g_{J'}$ & 1.031 (1.0334) & 1.17346 (1.17899) \\
LS purity (\%) & 99 / 100 & 94 / 97 \\
$S(M1)$ ($\mu_B^2$) & 10.9 & 25.6 \\
$\Gamma/2\pi$ (Hz) & 1.43 & 0.518 \\
$\mu_{\rm eff}$ ($\mu_B$) & 0.915 & 1.23 \\
\end{tabular}
\end{ruledtabular}
\end{table}

\subsection{Electric polarizability and ac Stark shift}
\label{sec:atomic_pol}
\label{sec:theory_E1}

The E1 response is governed by the dynamic scalar polarizability $\alpha_E(\omega)$ \cite{Mitroy2010}. We adopt $\alpha_E\simeq166$~a.u. for Er, measured at $1064.5$~nm \cite{Becher2018Er}, and the calculated value of $164$~a.u. for Dy \cite{Li2017Dy}. These values can be used at the M1 transition resonant wavelength $\lambda_0$ because $\alpha_E$ varies little in the infrared, far from the strong E1 resonances. For comparison, measurements for Er at $1570$~nm have reported $\alpha_E=163$~a.u. \cite{Becher2018Er}.

The broad lines that dominate $\alpha_E$ lie near $400$~nm, more than $500$~THz away from $\lambda_0$ \cite{Lepers2014Er,Becher2018Er,Li2017Dy}. In Er, the nearest odd-parity levels lie only $6.5$ and $22.1$~THz from $\lambda_0$ \cite{Kramida2024NIST}, but their hertz-level linewidths \cite{Patscheider2021Er} limit their contribution to less than $10^{-2}$~a.u., five orders of magnitude below $\alpha_E$; in Dy, the nearest levels lie more than $100$~THz away \cite{Kramida2024NIST,Petersen2019Dy}. Therefore, the E1 background varies smoothly around $\lambda_0$.

The vector and tensor parts are small, with a tensor-to-scalar ratio of about $3\%$ for Er \cite{Becher2018Er} and below $1\%$ for Dy \cite{Ravensbergen2018Dy,Li2017Dy}, so we use a single scalar $\alpha_E$ for all sublevels, with an uncertainty of a few percent. The corresponding ac Stark shift is the conventional optical dipole potential \cite{Grimm2000},
\begin{equation}
U_E(\bm r) = -\frac{|\bm E(\bm r)|^2}{4}\,\alpha_E(\omega),
\label{eq:UE}
\end{equation}
which is attractive for $\alpha_E>0$ and independent of $m_J$.

\section{Competition between the M1 and E1 potentials}
\label{sec:theory}
\label{sec:theory_saturation}

Because the magnetic light shift is dominated by a single M1 transition, each ground sublevel $m_J$ couples, for $\sigma^+$ light, only to the excited sublevel $m_J+1$ and forms an effective two-level system. The total potential felt by the atom is $U_{\rm tot}(\bm r,m_J)=U_M(\bm r,m_J)+U_E(\bm r)$, and the central question is under which conditions $U_M$ dominates.

For this two-level system, the saturation parameter is
\begin{equation}
s=\frac{\Omega_{m_J}^2/2}{\delta^2+\Gamma^2/4}\simeq\frac{\Omega_{m_J}^2}{2\delta^2}.
\label{eq:saturation}
\end{equation}

We restrict the calculations to a maximum saturation parameter $s_{\max}=0.01$. At this value, the second-order description of the light shift is accurate to $0.5\%$; the choice of $s_{\max}$ is justified in Sec.~\ref{sec:loss_single}.

At fixed $s_{\max}$, the relation $\Omega_{m_J}^2=2s_{\max}\delta^2$ links the field amplitude to the detuning. In the paraxial configurations considered below, the field amplitudes of each traveling wave obey $|\bm E|=c|\bm B|$. Substituting this relation into Eqs.~\eqref{eq:UM_master} and \eqref{eq:UE} gives
\begin{align}
U_M(\delta) &= a\,\delta, & a&=\frac{\hbar s_{\max}}{2}, \label{eq:UM_linear}\\
U_E(\delta) &= -b\,\delta^2, & b&= \frac{\alpha_Ec^2\hbar^2s_{\max}}{2\mu_{\rm eff}^2}. \label{eq:UE_quadratic}
\end{align}
Thus, at fixed saturation, the magnetic contribution grows linearly with $|\delta|$, whereas the electric background contribution grows quadratically. Equating the magnitudes of the two contributions defines the critical detuning,
\begin{equation}
\delta_c = \frac{a}{b} = \frac{\mu_{\rm eff}^2}{\alpha_Ec^2\hbar},
\label{eq:critical_detuning}
\end{equation}
at which $|U_M|=|U_E|$. Consequently,
\begin{equation}
\left|\frac{U_M}{U_E}\right| = \frac{\delta_c}{|\delta|},
\label{eq:UM_UE_ratio}
\end{equation}
so that the magnetic contribution dominates for $|\delta|<\delta_c$, while the electric background dominates for $|\delta|>\delta_c$. The saturation parameter fixes the field, and hence the power required at a given detuning, but $\delta_c$ is independent of $s$ and, therefore, independent of the power. For $m_J=-J$, $\delta_c/2\pi\simeq442$~MHz for Er and $805$~MHz for Dy.

Equation~\eqref{eq:critical_detuning} explains why the magnetic contribution can dominate despite its intrinsic weakness. Using $\mu_B=(\alpha_{\rm fs}/2)\,ea_0c$ and representing the E1 response as an effective line of dipole moment $d_{\rm eff}$ and detuning $\Delta_{\rm eff}$, with $\alpha_E=d_{\rm eff}^2/\hbar\Delta_{\rm eff}$, Eq.~\eqref{eq:UM_UE_ratio} becomes
\begin{equation}
\left|\frac{U_M}{U_E}\right|=\left(\frac{\alpha_{\rm fs}}{2}\right)^2\left(\frac{\mu_{\rm eff}/\mu_B}{d_{\rm eff}/ea_0}\right)^2\frac{\Delta_{\rm eff}}{|\delta|}.
\label{eq:ratio_atomic_units}
\end{equation}
The intrinsic suppression $(\alpha_{\rm fs}/2)^2\simeq1.3\times10^{-5}$ of the magnetic coupling is compensated by the ratio of detunings, $\Delta_{\rm eff}/|\delta|\sim10^6$, while the squared ratio of matrix elements is $0.10$ for Er and $0.18$ for Dy. Here $\Delta_{\rm eff}\simeq2\pi\times345$~THz for the dominant $401$~nm (Er) and $421$~nm (Dy) lines, including the counter-rotating term.

\begin{figure}[!htb]
\centering
\includegraphics[width=\columnwidth]{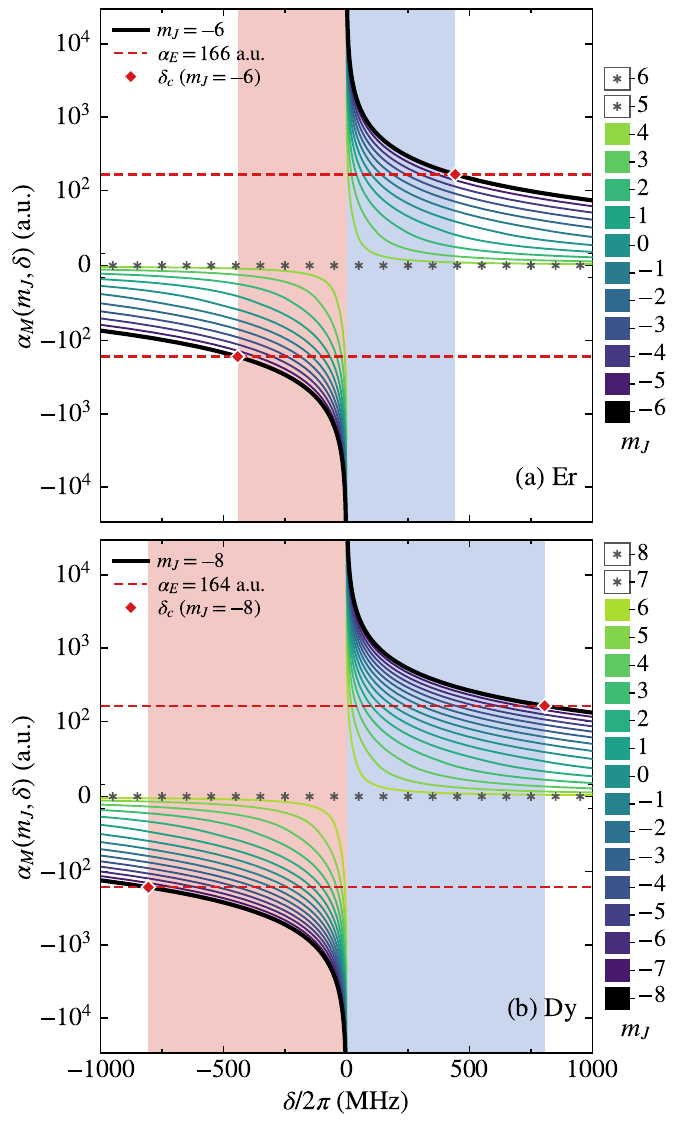}
\caption{Magnetic polarizability $\alpha_M(m_J,\delta)$ of the M1 transitions of (a) Er and (b) Dy for $\sigma^+$ light, on a symmetric logarithmic scale. The black curve is the reference state $m_J=-J$, the colored curves are the other sublevels, and the asterisks mark the $\sigma^+$-dark sublevels. Dashed lines mark $\pm\alpha_E$, red markers indicate $\delta_c$, and shaded regions indicate $|\delta|<\delta_c$. The M1 potential is attractive for $\delta<0$ and repulsive for $\delta>0$.}
\label{fig:polarizability}
\end{figure}

To compare both responses on the same scale, it is convenient to write the magnetic polarizability of Eq.~\eqref{eq:UM_master} in the units of $\alpha_E$,
\begin{equation}
\alpha_M(m_J)\equiv\frac{\alpha_v(\delta)\,r(m_J)}{c^2}=\frac{\mu_{\rm eff}(m_J)^2}{\hbar\delta c^2},
\label{eq:alphaM}
\end{equation}
so that $|\alpha_M|/\alpha_E=|U_M/U_E|$ for $|\bm E|=c|\bm B|$, and the critical detuning is the one at which $|\alpha_M(-J)|=\alpha_E$. Unlike $\alpha_E$, which does not change sign around the M1 transition, $\alpha_M$ carries the sign of $\delta$ and is negative where the potential is attractive. Figure~\ref{fig:polarizability} shows $\alpha_M$ across the Zeeman ladder.

\section{Trapping geometries}
\label{sec:schemes}

\begin{figure*}[t]
\centering
\includegraphics[width=\textwidth]{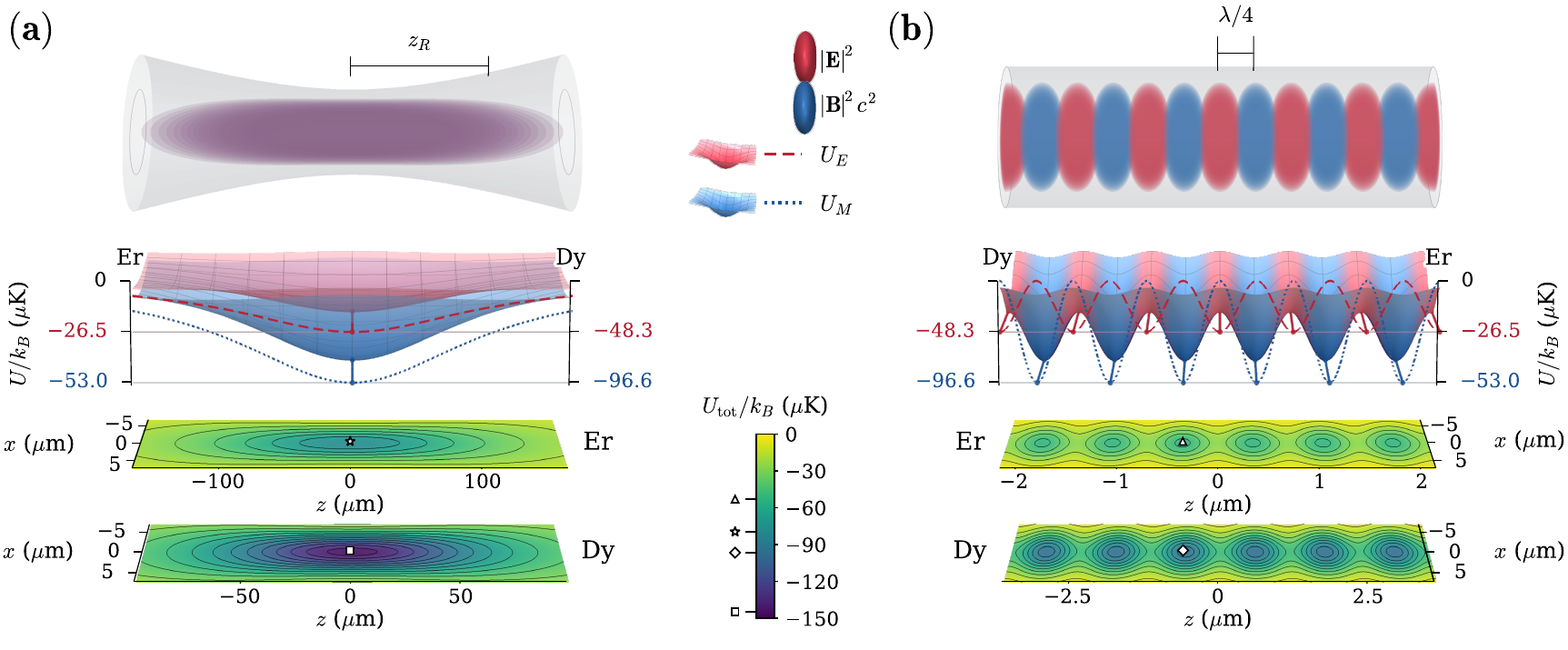}
\caption{Optical potentials for atoms in $m_J=-J$ at the operating point (Table~\ref{tab:operating}), for (a) a Gaussian tweezer and (b) the standing wave formed by two counterpropagating beams. Top: schematic of $|\bm E|^2$ and $c^2|\bm B|^2$. Middle: $U_E$ and $U_M$ along the beam axis; in these units the profiles are the same for both species, and the side axes give the Er and Dy scales. Bottom: $U_{\rm tot}$ in the $x$-$z$ plane for Er and Dy.}
\label{fig:trapping_potentials}
\end{figure*}

We propose and study two trapping configurations (Fig.~\ref{fig:trapping_potentials}). The first is an optical tweezer formed by a single circularly polarized Gaussian beam. The second is a one-dimensional lattice formed by two counterpropagating beams. In both scenarios, a bias field $\bm B_0$ along the beam axis defines the quantization axis.

For any given geometry and fixed $s_{\max}$, the light detuning is the only free parameter, and it is convenient to measure it in units of the critical detuning, $x=|\delta|/\delta_c$. From Eqs.~\eqref{eq:UM_linear}-\eqref{eq:UM_UE_ratio}, the magnetic contribution grows as $x$ and the electric contribution as $x^2$, with $|U_E|=x|U_M|$.

\subsection{Gaussian tweezer}
\label{sec:schemes_tweezer}

For a Gaussian beam propagating along $\hat{\bm z}$, the paraxial fields are
\begin{eqnarray}
\bm E(\bm r,t) &=& \frac{E_0}{\sqrt{2}}\,f(\rho,z)\,\hat{\bm e}(t), \label{eq:gaussian_field_E}\\
\bm B(\bm r,t) &=& \frac{\hat{\bm z}\times\bm E(\bm r,t)}{c}. \label{eq:gaussian_field_B}
\end{eqnarray}
The light polarization, $\hat{\bm e}(t)=\hat{\bm x}\cos\omega t+\hat{\bm y}\sin\omega t$, rotates in the transverse plane and drives $\sigma^+$ transitions. The Gaussian envelope is $f(\rho,z)=[w_0/w(z)]\,e^{-\rho^2/w^2(z)}$, where $\rho$ is the distance from the axis, $w(z)=w_0\sqrt{1+z^2/z_R^2}$, and $z_R=\pi w_0^2/\lambda$ is the Rayleigh length. The two potentials follow the same envelope and add at the focus [Fig.~\ref{fig:trapping_potentials}(a)], where the depth is
\begin{equation}
U_{\rm tw}=|U_M|+|U_E|=\frac{\hbar s_{\max}\delta_c}{2}\,x(1+x).
\label{eq:tweezer_depth}
\end{equation}
Because $z_R\gg w_0$ in the paraxial regime, the axial confinement is much weaker than the radial one (Table~\ref{tab:operating}).

\subsection{Standing-wave lattice}
\label{sec:schemes_lattice}

If the fields of the two counterpropagating beams rotate in the same direction about $\bm B_0$, the standing wave drives $\sigma^+$ transitions at every point. Considering each beam as in Eqs.~\eqref{eq:gaussian_field_E} and \eqref{eq:gaussian_field_B}, the fields of the standing wave are
\begin{align}
\bm E(\bm r,t) &= \sqrt{2}E_0\,f(\rho,z)\cos(kz)\,\hat{\bm e}(t), \label{eq:lattice_E}\\
\bm B(\bm r,t) &= \frac{\sqrt{2}E_0}{c}\,f(\rho,z)\sin(kz)\,\hat{\bm e}(t). \label{eq:lattice_B}
\end{align}
The magnetic antinodes, at $z=(2n+1)\lambda/4$, coincide with the electric nodes and vice versa. The lattice thus contains two sublattices of period $\lambda/2$, displaced by $\lambda/4$ [Fig.~\ref{fig:trapping_potentials}(b)]: one of magnetic wells, where $\bm E=0$, and one of electric wells, where $\bm B=0$. The potential barrier between neighboring sites is
\begin{equation}
\Delta U_{\rm lat}=|U_M|-|U_E|=\frac{\hbar s_{\max}\delta_c}{2}\,x(1-x),
\label{eq:lattice_modulation}
\end{equation}
which is maximal at $x=1/2$, where $|U_M|=2|U_E|$.

To compare both trapping geometries, we fix the Gaussian beam waist at $w_0=7~\mu$m, the saturation parameter at $s_{\max}=0.01$, and the detuning at $x=1/2$. Table~\ref{tab:operating} summarizes the trapping parameters for both geometric configurations for Er and Dy atoms. Figure \ref{fig:trapping_potentials} shows the electric-only, magnetic-only, and total trapping potentials for both configurations.

\begin{table}[!htb]
\caption{Operating point of Fig.~\ref{fig:trapping_potentials} ($x=1/2$, $s_{\max}=0.01$, $w_0=7~\mu$m). The detuning $\delta$ and the optical magnetic field at the trapping site, $B_{\rm site}$, are common to both geometries. The depth is $U_{\rm tw}$ [Eq.~\eqref{eq:tweezer_depth}] in the tweezer and $|U_M|$ at the magnetic lattice sites, and $\Delta U_{\rm lat}$ [Eq.~\eqref{eq:lattice_modulation}] is the barrier between neighboring sites. $\nu_r$ and $\nu_z$ are the radial and axial trap frequencies for $^{166}$Er and $^{164}$Dy, and $P$ is the power per beam.}
\label{tab:operating}
\begin{ruledtabular}
\begin{tabular}{lcccc}
 & \multicolumn{2}{c}{Er} & \multicolumn{2}{c}{Dy} \\
\colrule
$\delta/2\pi$ (MHz) & \multicolumn{2}{c}{$-221$} & \multicolumn{2}{c}{$-402$} \\
$B_{\rm site}$ (G) & \multicolumn{2}{c}{24.4} & \multicolumn{2}{c}{33.1} \\
\colrule
 & Tweezer & Lattice & Tweezer & Lattice \\
\colrule
Depth$/k_B$ ($\mu$K) & 79.5 & 53.0 & 144.9 & 96.6 \\
$\Delta U_{\rm lat}/k_B$ ($\mu$K) & & 26.5 & & 48.3 \\
$\nu_r$ (kHz) & 2.9 & 2.3 & 3.9 & 3.2 \\
$\nu_z$ (kHz) & 0.13 & 35.9 & 0.30 & 28.9 \\
$P$ (mW) & 55 & 13.7 & 101 & 25.2 \\
\end{tabular}
\end{ruledtabular}
\end{table}

\section{Trap lifetime and practical requirements}
\label{sec:loss_mechanisms}

The M1 trapping schemes are subject to the same loss mechanisms as conventional E1 dipole traps; however, the long lifetime of the excited state changes their relative importance. Figure~\ref{fig:timescales} shows the time scales of each process as a function of the saturation parameter.

\subsection{Photon scattering}
\label{sec:loss_single}

\paragraph{M1 scattering.} At low saturation, the atom scatters photons at the rate $\Gamma_{\rm sc}=\Gamma s/2$. This scattering is elastic: each emission returns the atom to the trapped state, with a recoil heating below $1$~nK\,s$^{-1}$ \cite{Grimm2000}. However, the light also mixes the excited state with the ground state,
\begin{equation}
|\tilde e\rangle\simeq|e\rangle-\frac{\Omega_{m_J}}{2\delta}|g\rangle,
\label{eq:dressed_ground}
\end{equation}
so a fraction $(\Omega_{m_J}/2\delta)^2=s/2$ of the emissions leaves the atom in $|\tilde e\rangle$ \cite{Dalibard1985}. This state has a light shift of opposite sign and lives for $1/\Gamma\simeq0.11$~s (Er) and $0.3$~s (Dy), much longer than a trap period, so the atom is ejected at the rate $\Gamma_{\rm ej}=\Gamma_{\rm sc}\,s/2=\Gamma s^2/4$, quadratic in $s$ because it is the scattering rate weighted by the probability that an emission leaves the atom in $|\tilde e\rangle$. In conventional dipole traps, in contrast, the excited state decays within tens of nanoseconds and this process does not eject the atom.

\begin{figure}[!htb]
\centering
\includegraphics[width=\columnwidth]{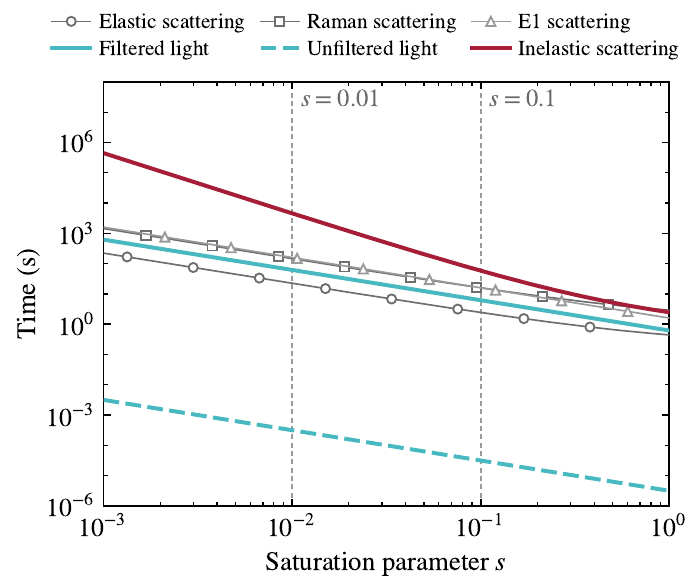}
\caption{Time scales of the light-induced processes as a function of the saturation parameter, for an Er atom in $m_J=-J$ at $|\delta|=\delta_c/2$. The resonant light corresponds to a laser with $\gamma_L=100$~kHz, unfiltered and filtered by a $1$~MHz-wide filter, and E1 scattering is the upper bound for the tweezer. We show only Er because it is the most demanding case: for Dy the curves have the same shape but are shifted to longer times, mainly because of its narrower M1 line and larger detuning. Table~\ref{tab:timescales} lists the values for both species at the operating point.}
\label{fig:timescales}
\end{figure}

\begin{table}[!htb]
\caption{Time scales of Fig.~\ref{fig:timescales} at the operating point ($s=0.01$, $|\delta|=\delta_c/2$), with the light filtered by a $1$~MHz-wide filter.}
\label{tab:timescales}
\begin{ruledtabular}
\begin{tabular}{lcc}
Process & Er (s) & Dy (s) \\
\colrule
Elastic scattering & 22 & 62 \\
Depolarization (Raman) & 146 & 527 \\
E1 scattering & $>160$ & $>320$ \\
Filtered resonant light & 62 & 206 \\
Inelastic scattering & $4.6\times10^3$ & $1.3\times10^4$ \\
\end{tabular}
\end{ruledtabular}
\end{table}

\paragraph{Raman scattering.} The photon reemitted from $m_J'=-J+1$ can also be $\pi$ or $\sigma^-$, in a fraction $2/(2J+1)$ of the events, leaving the atom in a neighboring sublevel. The atom remains trapped, but the stretched state depolarizes in a time $\tau_{\rm dep}=(2J+1)/(2\Gamma_{\rm sc})$.

\paragraph{E1 scattering.} This contribution is weaker because it originates from broad lines near $400$~nm. Attributing all of $U_E$ to the nearest transition, at $401$ nm (Er) or $421$ nm (Dy), gives an upper bound for the tweezer \cite{Grimm2000,Frisch2012ErMOT,Lu2010DyMOT}. This contribution is further reduced in the lattice configuration, where $\bm E=0$ at the center of the magnetic lattice site.

\paragraph{Residual resonant light.}
\label{sec:loss_noise}
Lasers emit a weak spectral background that extends far from the carrier, caused by frequency noise, amplified spontaneous emission, or modulation sidebands. This background is harmful despite the weakness of the M1 transition because, in a near-resonant trap, the light that falls on the resonance cannot be neglected.

Considering a laser with a Lorentzian spectrum of full width at half maximum $\gamma_L$ \cite{DiDomenico2010}, we integrate over the atomic absorption cross section \cite{Foot2005} to obtain the excitation rate $R=\pi s\,\gamma_L$.

In contrast to E1 trapping configurations, which use large detuning and extremely low saturation parameters, magnetic trapping with light requires non-negligible saturation and a near-resonant condition, because the M1 potential dominates only for $|\delta|<\delta_c$.

Because lifetimes of $\sim1$ s already require $\gamma_L\lesssim30$~Hz, it is convenient to suppress the background noise beyond standard means of laser narrowing. For example, for $\gamma_L=100$~kHz, an external cavity acting as a $1$~MHz filter attenuates the on-resonance noise level by $5\times10^{-6}$ (Er) and $1.5\times10^{-6}$ (Dy), extending the lifetime to $62$~s for Er and $206$~s for Dy (Fig.~\ref{fig:timescales} and Table~\ref{tab:timescales}). 

\paragraph{Choice of the saturation parameter.} The depth, power, and almost all scattering rates grow as $s$, but inelastic scattering grows as $s^2$ (Fig.~\ref{fig:timescales}). At $s=0.1$ the trap would be ten times deeper, but inelastic scattering would remain only one order of magnitude below the loss by resonant light; at $s=0.01$ it remains almost two orders below. At this point (Table~\ref{tab:timescales}), the filtered resonant light sets the lifetime, longer than an experimental cycle, and Raman scattering limits the spin polarization. Losses from collisions with the background gas depend on the vacuum of each apparatus and are not included. For reference, single Dy atoms have been held in tweezers for $300$~s \cite{Bloch2023DyTweezer}.

\subsection{Collisions and bias field}
\label{sec:loss_collisions}
\label{sec:loss_dipolar}

\paragraph{Dipolar relaxation.} Atoms in $m_J=-J$, the lowest Zeeman sublevel, cannot release energy by flipping their spin; therefore, several atoms can share a trapping site \cite{Lahaye2009}. An atom in another sublevel, left by imperfect preparation or by Raman scattering, reopens the channel \cite{Hensler2003}: each spin flip releases $g_J\mu_BB_0$, approximately $12~\mu$K per atom at the $0.3$~G used to load Er into dipole traps \cite{Aikawa2012ErBEC}. Because the atoms that change sublevels remain at their site, a site holding $N$ atoms loses its polarization after approximately $\tau_{\rm dep}/N$. Once an atom has left the stretched state, the subsequent collisions heat the atoms until the site holds zero or one atom. Singly occupied sites are unaffected.

\paragraph{Bias field.} The field $\bm B_0$ must avoid the dense low-field Feshbach resonances of Er and Dy \cite{Frisch2014ErChaos,Baumann2014DyFeshbach}, be aligned with the beam, and exceed the fictitious field of the vector light shift, approximately $0.06$~G (Er) and $0.07$~G (Dy). The Zeeman shift produced on the transition, $2.6$~MHz/G (Er) and $2.4$~MHz/G (Dy), is absorbed by readjusting $\delta$.

\section{Discussion}
\label{sec:discussion}

\paragraph{Feasibility of magnetic trapping.} An optical M1 transition can dominate the trapping of a neutral atom. In Er and Dy, the M1 interaction exceeds the E1 background over the entire range $|\delta|<\delta_c$, which extends for several hundred MHz independently of the optical power (Table~\ref{tab:operating}). The atoms can remain trapped for about a minute (Er) and several minutes (Dy), and retain their spin polarization for a comparable time (Table~\ref{tab:timescales}).

\begin{figure}[!htb]
\centering
\includegraphics[width=\columnwidth]{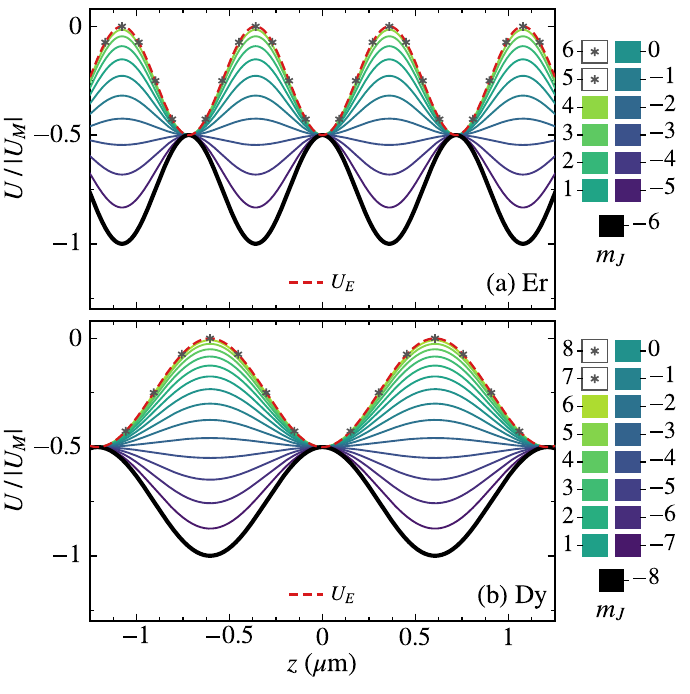}
\caption{State selectivity of the magnetic lattice for (a) Er and (b) Dy over the same range of $z$, at the operating point of Fig.~\ref{fig:trapping_potentials}(b). Each curve is the potential of one sublevel (key at right, $m_J=-J$ in black) in units of $|U_M|$ (Table~\ref{tab:operating}), with the Zeeman shift of the bias field neglected. The dashed red line is the E1 background, common to all sublevels and followed exactly by the $\sigma^+$-dark sublevels $m_J=J-1,J$ (asterisks).}
\label{fig:selectivity}
\end{figure}

\paragraph{State-dependent lattices.} The advantage of optical trapping in the M1 regime lies in the possibility of state-dependent atomic manipulation. In the proposed standing-wave configuration, each sublevel is trapped on one of two sublattices displaced by $\lambda/4$. At the electric sites, only the E1 background acts, with the same depth $|U_E|$ for all sublevels, whereas at the magnetic sites, the M1 interaction acts, with a depth $r(m_J)|U_M|$ that decreases from the stretched state. Each sublevel occupies the deepest trapping site available: the magnetic sublattice if $r(m_J)>x$ and the electric sublattice otherwise (Fig.~\ref{fig:selectivity}), where $x=|\delta|/\delta_c=|U_E|/|U_M|$ is the detuning in units of critical detuning. Increasing the detuning until $x>(J-1)/J$ leaves only the stretched state on the magnetic sublattice, at the cost of a lower barrier height. Switching the polarization from $\sigma^+$ to $\sigma^-$ also reverses the order of the sublevels, so that the stretched state moves to the electric sublattice. Thus, the stretched state is spatially separated from the other sublevels, which is the property exploited by state-dependent lattices for controlled collisions \cite{Mandel2003Collisions} and for the independent control of two internal states \cite{Heinz2020Sr}. The sublattice occupied by each sublevel can be verified through its trap frequency, which on the magnetic sublattice depends on $m_J$. In alkali atoms, this selectivity relies on the vector light shift of light tuned a few THz from broad lines \cite{Mandel2003Transport}, which scatters many more photons at an equal trapping depth.

\paragraph{Uncertainties of the calculation.} The main uncertainty is the M1 line strength, calculated in LS coupling and not yet measured for atoms in free space. For the analogous $^2F_{7/2}\rightarrow{}^2F_{5/2}$ transition of Tm, whose $4f$ shell lacks a single electron, so that its two levels are the only ones of the configuration and LS coupling is exact for them, Eq.~\eqref{eq:S_closed} predicts $\Gamma/2\pi\simeq1.7$~Hz in contrast with the measured bound $\Gamma/2\pi\le1.4$~Hz \cite{Sukachev2016Tm}, suggesting that LS coupling may overestimate $S(M1)$ by $15\%$ or more. For Dy, the LS purities alone suggest a reduction of order $10\%$. For Er, there is, in addition, an indirect test: the ${}^3H_5\rightarrow{}^3H_6$ emission of Er atoms isolated in solid neon has a measured lifetime of $78.9$~ms \cite{Zaitsev2026}. Since the refractive index $n$ of the host enhances M1 emission by $n^3$ and nonradiative processes can only shorten the lifetime \cite{Zaitsev2026}, the measurement implies $\Gamma/2\pi\lesssim2.0$~Hz in free space, consistent with the LS value. This uncertainty, together with that of $\alpha_E$, rescales $\delta_c$, the depths, and the optical powers by a comparable fraction, but does not compromise the trapping schemes, which only require $|\delta|<\delta_c$. Moreover, because the atoms are held by the M1 transition itself, the trap provides direct access to it: the number of trapped atoms or the trap frequency, measured as a function of the laser frequency, depend on the M1 coupling and would allow $S(M1)$ to be determined without resolving the hertz-wide line. The E1 field also shifts the excited level, but by nearly the same amount as the ground level, because both share the $6s$ electrons that dominate the E1 response. The differential polarizability of the analogous Tm transition has been measured to be only $-0.047(18)$~a.u. \cite{Golovizin2019Tm}. Even a $1\%$ difference between the two polarizabilities would displace the resonance by at most $10$~kHz, and at the magnetic lattice sites, where $\bm E=0$, this shift vanishes.

\paragraph{Other lanthanides.} According to Eq.~\eqref{eq:ratio_atomic_units}, the M1 interaction compensates for its intrinsic weakness through the ratio $\Delta_{\rm eff}/|\delta|$, so the scheme applies to any atom with an allowed M1 transition with distant E1 lines. Tm and Ho meet this condition: their M1 lines $^2F_{7/2}\rightarrow{}^2F_{5/2}$ at $1.14~\mu$m \cite{Sukachev2016Tm,Golovizin2019Tm} and $^4I_{15/2}\rightarrow{}^4I_{13/2}$ at $1845$~nm \cite{Kramida2024NIST} lie far from their strong E1 lines at $410.6$ and $410.5$~nm, used for laser cooling \cite{Sukachev2010TmMOT,Miao2014HoMOT}. The only stable isotope of each has a nonzero nuclear spin; therefore, their M1 response must include the hyperfine structure.

\section{Conclusions}
\label{sec:conclusions}

We have shown that the magnetic field component of light can dominate the optical trapping of neutral lanthanide atoms near their intraconfigurational M1 transitions, for Er at 1437 nm and Dy at 2419 nm. The ac Zeeman shift surpasses the ac Stark shift below a critical detuning, enabling optical potentials of order $100~\mu$K that maintain atomic spin polarization and trapping stability for one to several minutes. Although the M1 line strengths are LS-coupling calculations pending experimental verification, the M1 trapping regime offers a pathway to measure these parameters. The approach extends to other lanthanides with infrared M1 transitions, such as Tm and Ho, if their hyperfine structure is included. The distinctive feature of M1 trapping is its state-dependent lattice structure, where atoms are confined at nodes of the electric field, so that Zeeman sublevels occupy magnetic or electric sublattices according to their coupling. This opens new avenues for the quantum control of lanthanide atoms beyond electric-dipole trapping.

\begin{acknowledgments}
P.S., E.O., and N.V. acknowledge support from the National Agency for Research and Development (ANID) through Project FONDECYT Grants No. 1240204, ANID-Subdirección de Capital Humano/Magíster Nacional/2025-22251958 and ANID-Subdirección de Capital Humano/Doctorado Nacional/2022-21221251, respectively.
\end{acknowledgments}

\bibliography{references}
\end{document}